\documentstyle[psfig,12pt,a4wide]{article}
\newcommand{\news}{\setcounter{equation}{0}}

\newcommand{\grad}{\mbox{$\bigtriangledown$}}
\def\eqn{\begin{equation}} 
\def\eeqn{\end{equation}}
\def\arr{\begin{array}} 
\def\earr{\end{array}}
\def\eqna{\begin{eqnarray}} 
\def\eeqna{\end{eqnarray}} 
\def\a{\alpha}
 
\def\D{\Delta}
\def\s{\sigma}

\def\o{\omega}
\def\r{\rho}
\def\O{\Omega}
\def\P{\Psi}
\def\e{\epsilon}
\def\th{\theta} 
\def\m{\mu} 
\def\n{\nu} 
 
\def\z{\zeta}
 
\def\p{\partial} 
\def\g{\gamma} 
\font\mybb=msbm10 at 12pt 
\def\bb#1{\hbox{\mybb#1}}

\begin{document}

\vspace*{-.6in}
\thispagestyle{empty}
\begin{flushright}
PUPT-1906\\
\end{flushright}

\vspace{.3in}
{\Large 
\begin{center}
{\bf Absorption by Double-Centered D3-Branes and \\
the Coulomb Branch of ${\cal N}=4$ SYM Theory}
\end{center}}
\vspace{.3in}
\begin{center}
Miguel S. Costa\footnote{miguel@feynman.princeton.edu}\\
\vspace{.1in}
\emph{Joseph Henry Laboratories\\ Princeton University \\ Princeton, New Jersey 08544, USA}
\end{center}

\vspace{.5in}

\begin{abstract}
We calculate the classical cross-section for absorption of a minimally coupled scalar in the 
double-centered D3-brane geometry. The dual field theory has gauge symmetry broken to 
$S(U(N_1)\times U(N_2))$ and is on the Coulomb branch of ${\cal N}=4$ Super Yang-Mills 
theory. Our analysis is valid at energy scales much smaller than the W particles mass, giving 
logarithmic corrections to the cross section calculated at the IR conformal fixed points. 
These corrections are associated with deformations of the ${\cal N}=4$ Super Yang-Mills theory 
by irrelevant operators that break conformal invariance and correspond to processes where 
a virtual pair of gauge particles or a virtual pair of W bosons interact with the 
incident wave to create a pair of gauge particles. 
\end{abstract}
\newpage

\section{Introduction}
\news

The famous AdS/CFT duality \cite{Mald,GKP,Witten} 
has been investigated intensively over the past two years
(see \cite{review} for a review and an exhaustive list of references).
The most studied case is the conjectured equivalence between type IIB string theory on
$AdS_5\times S^5$ and the low energy theory on $N$ D3-branes given by ${\cal N}=4$ Super
Yang-Mills (SYM) theory with $SU(N)$ gauge group. Unfortunately, due to our limited 
knowledge of Ramond-Ramond backgrounds in string theory, we are bound to consider
type IIB supergravity on $AdS_5\times S^5$,
which is dual to large $N$ SYM theory at strong coupling. 

The moduli space of $SU(N)$ ${\cal N}=4$ SYM theory is $\bb{R}^{6(N-1)}/S_N$, 
parameterizing the relative position of $N$ identical D3-branes in the transverse space
$\bb{E}^6$. At the origin of the moduli space the theory is superconformally
invariant. This fact is related to symmetry enhancement at the D3-brane near-horizon
geometry \cite{GibbTown} and it guarantees that both theories are invariant under the 
$PSU(2,2|4)$ supergroup. However, many interesting theories like QCD are not conformally 
invariant. The main purpose of this paper is to study the AdS/CFT duality when conformal 
invariance is broken by separating the D3-branes. In particular, we consider the case 
of two parallel stacks of D3-branes. In the dual field theory, 
moving away from the origin of the moduli space means that the scalar fields have 
non-zero expectation values. This is also known as the Coulomb branch of the theory.
The dual gravitational configuration is related to the double-centered type IIB 
D3-brane background. Previous work on the Coulomb branch of the correspondence 
can be found in $[6-17]$. Other studies of the duality with broken conformal 
invariance are $[18-27]$.

Consider $N=N_1+N_2$ D3-branes described at low energies by the $SU(N)$ ${\cal N}=4$ SYM theory.
Next, Higgs the system by giving an expectation value to the scalar fields according to
\eqn
\frac{\langle\vec{Y}\rangle}{2\pi\a'}\equiv 
\langle \vec{\phi} \rangle =\frac{\vec{\D}}{2\pi\a'}
\left(
\arr{cc}
I_1 & 0 \\
0 & -I_2
\earr
\right)\ ,
\label{Higgs}
\eeqn
where $I_i$ is the $N_i\times N_i$ unit matrix. The $N_1$ and $N_2$ D3-branes are
separated by a distance $2\D$ and the gauge symmetry is broken 
to $S(U(N_1)\times U(N_2))$. Furthermore, the theory is no longer conformally invariant 
and there is a scale naturally set by the W particles mass that arises from the symmetry breaking
process:
\eqn
m_W=\frac{\D}{\pi\a'}\ .
\label{wmass}
\eeqn
The theory is still maximally supersymmetric, i.e. it has 16 conserved supercharges (but
it is not invariant under the extra 16 superconformal charges). Hence, we shall be interested
in maximally supersymmetric RG flows \cite{Intr}. In the UV, for energies $\e\gg m_W$,
we recover the conformal invariant $SU(N)$ theory, while for $\e\ll m_W$ the theory flows
to the IR conformal fixed points with $SU(N_i)$ gauge symmetry. In particular, the gauge
coupling beta function vanishes along such flows and the theory can be described starting
from a IR fixed point by a deformation given by a non-renormalizable, irrelevant operator
with scaling dimension 8, as was conjectured in \cite{GHKK,GubserHash,Intr}.

We shall be able to test the above conjecture by studying the low energy scattering
of the dilaton field (or more generally of a minimally coupled scalar) in the dual double-centered
D3-brane geometry. To be more precise, we find logarithmic corrections to the cross-section 
calculated at the IR conformal fixed points. These corrections are associated with processes 
where a pair of gauge particles is created via the interaction of the incident dilaton wave
with a virtual pair of gauge particles or a virtual pair of W bosons. Hence, even at energies
$\e\ll m_W$, well below the threshold for W boson pair production, the dual gravity theory 
`knows' about the W particles. This result provides further evidence for the duality to hold
even when conformal invariance is broken, as is the case in the Coulomb branch of the
field theory away from the origin of the moduli space of vacua. 

The paper is organized as follows: In Section 2 we introduce the prolate spherical coordinates,
which are very useful to solve scattering problems in a double-centered black hole background
\cite{Chand}. Then, we analyze the double-centered 
D3-brane geometry in the decoupling limit. In Section
3 we calculate the classical cross-section for absorption of a minimally coupled scalar in 
the double-centered D3-brane background by using the improved matching
technique developed in \cite{GHKK}. These results are interpreted from the dual field theory 
point of view in Section 4. Finally, we give some 
concluding remarks in Section 5. 

\section{Prolate Spherical Coordinates and Decoupling Limit}
\news

In this section we analyze the double-centered D3-brane geometry and the corresponding decoupling 
limit using prolate spherical coordinates \cite{Chand}. The D3-brane geometry is described by 
the metric 
\eqn
ds^2=H^{-1/2}ds^2(\bb{M}^4)+H^{1/2}ds^2(\bb{E}^6)\ .
\label{D3-metric}
\eeqn
In the double-centered case the harmonic function $H$ is given by
\eqn
H=1+\frac{R_1^{\ 4}}{|\vec{y}-\vec{\D}|^4}
+\frac{R_2^{\ 4}}{|\vec{y}+\vec{\D}|^4}\ ,
\label{H}
\eeqn
where $\vec{y}$ are the coordinates in the space orthogonal to the branes $\bb{E}^6$ and
\eqn
R_i^{\ 4}=4\pi\a'^2g_sN_i\ .
\label{R}
\eeqn
This solution is associated with a system of $N_1$ D3-branes placed at $\vec{\D}$ and
$N_2$ D3-branes at $-\vec{\D}$.

Next, we place the branes along the $y^6$ axis and change to prolate spherical 
coordinates
\eqn
\arr{rcl}
y_6=r\cos{\th_1}&\equiv&\D\cosh{\P}\cos{\th}\ ,\\
\r=r\sin{\th_1}&\equiv&\D\sinh{\P}\sin{\th}\ ,
\earr
\label{psc}
\eeqn
where $(r,\th_1,\th_2,...,\th_5)$ are the usual spherical coordinates and 
$(\P,\th,\th_2,...,\th_5)$ are the prolate spherical coordinates. Of course, the system has
$SO(5)$ invariance and the transformation $(r,\th_1)\rightarrow(\P,\th)$ is given by
\eqn
\arr{rcl}
r^2&\equiv&\D^2\left( \cosh^2{\P}\cos^2{\th}+\sinh^2{\P}\sin^2{\th}\right)\ ,\\
\tan{\th_1}&\equiv&\tanh{\P}\tan{\th}\ .
\earr
\label{psc2}
\eeqn
It is convenient to define the $(\eta,\mu)$ coordinates by
\eqn
\eta=\cosh\P\ ,\ \ \ \mu=\cos\th\ ,
\label{psc3}
\eeqn
with $\eta\in ]1,+\infty[$ and $\mu\in[-1,1]$. For $\eta\gg 1$, we have $\eta\sim r\D$ and
$\th\sim\th_1$, i.e. the spherical coordinates are recovered. The coordinate system is singular 
at $\eta=1$ (and $\mu$ arbitrary). In prolate spherical coordinates 
the harmonic function $H$ becomes
\eqn 
H=1+\frac{R_1^{\ 4}}{\D^4(\eta-\mu)^4}+\frac{R_2^{\ 4}}{\D^4(\eta+\mu)^4}\ ,
\label{H2}
\eeqn
and the metric element on the transverse space $\bb{E}^6$
\eqn
ds^2(\bb{E}^6)=\D^2
\left[(\eta^2-\mu^2)\left(\frac{d\eta^2}{\eta^2-1}+\frac{d\mu^2}{1-\mu^2}\right)
+(\eta^2-1)(1-\mu^2)d\O_4^{\ 2}\right]\ .
\label{transvmetric}
\eeqn
In the limit $\eta\gg 1$ we have the asymptotics of the $N=N_1+N_2$ charged D3-brane solution.

We proceed by analyzing the geometry down the throats. Consider the limit 
$\eta\rightarrow 1$ and $\mu\rightarrow 1$ (or $-1$). This corresponds to the geometry near the
horizon of the first (second) stack of $N_1$ ($N_2$) branes. It is convenient to perform
the following change of coordinates
\eqn
\eta=1+a\ ,\ \ \ \mu=\pm 1\mp b\ .
\label{throatcoord}
\eeqn
Next, define the radial coordinate from the center of the $i$-th throat $\z_i$ and the angle 
with the $\mu=\pm 1$ direction $\varphi_i$ by 
\eqn
\z_i=a+b\ ,\ \ \ \cos{\varphi_i}=\frac{a-b}{a+b}\ .
\label{throatcoord2}
\eeqn
To leading order in $a$ and $b$ (i.e. $\z_i$) the spacetime metric reads
\eqn
ds^2=H^{-1/2}ds^2(\bb{M}^4)+H^{1/2}\left( d\z_i^{\ 2} + \z_i^{\ 2}d\O_5^{\ 2}\right)\ ,
\label{throatmetric}
\eeqn
with
\eqn
H=1+\left(\frac{R_j}{2\D}\right)^4+\left(\frac{R_i}{\D\z_i}\right)^4\ .
\label{H3}
\eeqn
Note that here and henceforth we have $i\ne j$.
Of course, for $\z_i\ll 1$ far down the throat the constant terms in the harmonic 
function $H$ become irrelevant.
Then, the geometry becomes the well known $AdS_5\times S^5$ space with radius $R_i$.

To analyze the geometry in the decoupling limit, first consider the double-centered
D3-brane solution. This is an asymptotically flat space with total charge $N=N_1+N_2$
and with two throats centered at $\pm\vec{\D}$ (see figure 1(a)). In the previous
paragraph we saw that the throat geometries are spherically symmetric and that 
far down the throats
we obtain the $AdS_5\times S^5$ single-centered black hole near-horizon geometry. The 
decoupling limit is defined by sending $\a'\rightarrow 0$ such that
\eqn
\vec{U}\equiv\frac{\vec{y}}{\a'}\ ,\ \ \ \vec{D}\equiv\frac{\vec{\D}}{\a'}\ ,
\label{declimit}
\eeqn
are kept fixed, as well as the dimensionless parameters $g_s$ and $N_i$ (we require
$g_s\ll 1$ and $g_sN_i\gg 1$ for the supergravity approximation 
to hold). The spacetime metric becomes
\eqn
\arr{rcl}
\displaystyle{\frac{ds^2}{\a'}}&=&
\displaystyle{\frac{1}{\sqrt{4\pi g_s}}
\left(\frac{N_1}{D^4(\eta-\mu)^4}+\frac{N_2}{D^4(\eta+\mu)^4}\right)^{-\frac{1}{2}}
ds^2(\bb{M}^4)}\\
&&+\displaystyle{\sqrt{4\pi g_s}
\left(\frac{N_1}{D^4(\eta-\mu)^4}+\frac{N_2}{D^4(\eta+\mu)^4}\right)^{\frac{1}{2}}
ds^2(\bb{E}^6)}\ ,
\earr
\label{dlm}
\eeqn
where $ds^2(\bb{E}^6)$ is given by (\ref{transvmetric}) with $\D^2$ replaced by $D^2$.
As we take the decoupling limit the throats' separation goes to zero and we have 
$R_i\gg\D$ (see figure 1(b)). 
In the limit we obtain the tree-like geometry sketched in figure 1(c) \cite{MichStro}.
This geometry is dual to the Coulomb branch of ${\cal N}=4$ SYM theory as described in
the Introduction. In the limit $\eta\gg 1$, the geometry approaches $AdS_5\times S^5$ near its
boundary with radius $R=R_1+R_2$. This corresponds to the conformal UV limit of the $SU(N)$
field theory. In the other limits $\eta\rightarrow 1$, $\mu\rightarrow \pm 1$ we obtain the 
$AdS_5\times S^5$ geometry with radius $R_i$. In the field theory side these correspond to the
$SU(N_i)$ IR conformal fixed points. At low energies 
$\e\ll m_W$, we can start from the IR theory and consider some deformation of the 
field theory. This means that the geometry (\ref{dlm}) is described by the throat metric
(\ref{throatmetric}) with the harmonic function $H$ given by
\eqn
H=\left(\frac{R_j}{2\D}\right)^4+\left(\frac{R_i}{\D\z_i}\right)^4\ .
\label{H4}
\eeqn
The constant term is due to the other throat and will be interpreted in the field theory 
side as a result of integrating out the massive W particles. 

A different possibility is to take the decoupling limit (\ref{declimit}) with $\D$ 
finite and non-zero. This means that $R_i\ll\D$ and that in the limit  
we have $\D/\a'\rightarrow \infty$. In this case, as we take the decoupling
limit both throats are kept apart (figure 1(d)) and in the limit we obtain two disconnected
$AdS_5\times S^5$ spaces (figure 1(e)). In the low energy limit $\e\ll m_W$, we shall study
the effects in the dual field theory that correspond to connecting the two throats to 
asymptotically flat space as described by figure 1(d).

\begin{figure}
\centerline{\psfig{figure=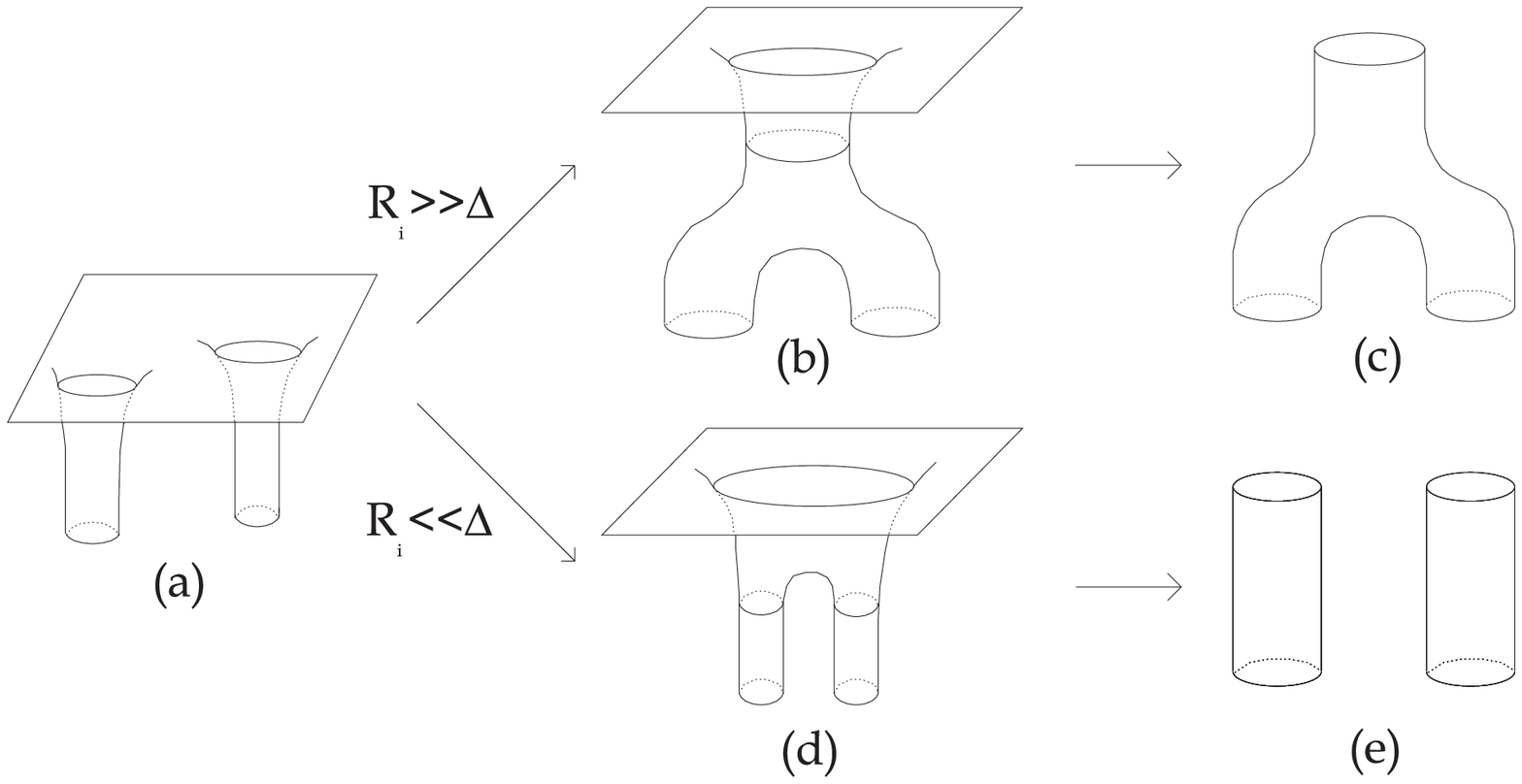,width=6in}}
\caption{\small{The different decoupling limits. The case (a)$\rightarrow$(b)$\rightarrow$(c)
corresponds to keeping $\D/\a'$ fixed,
while the case (a)$\rightarrow$(d)$\rightarrow$(e) to keeping $\D$ finite.}}
\label{fig1}
\end{figure} 

\section{Scattering of Minimally Coupled Scalar}
\news

\begin{figure}
\centerline{\psfig{figure=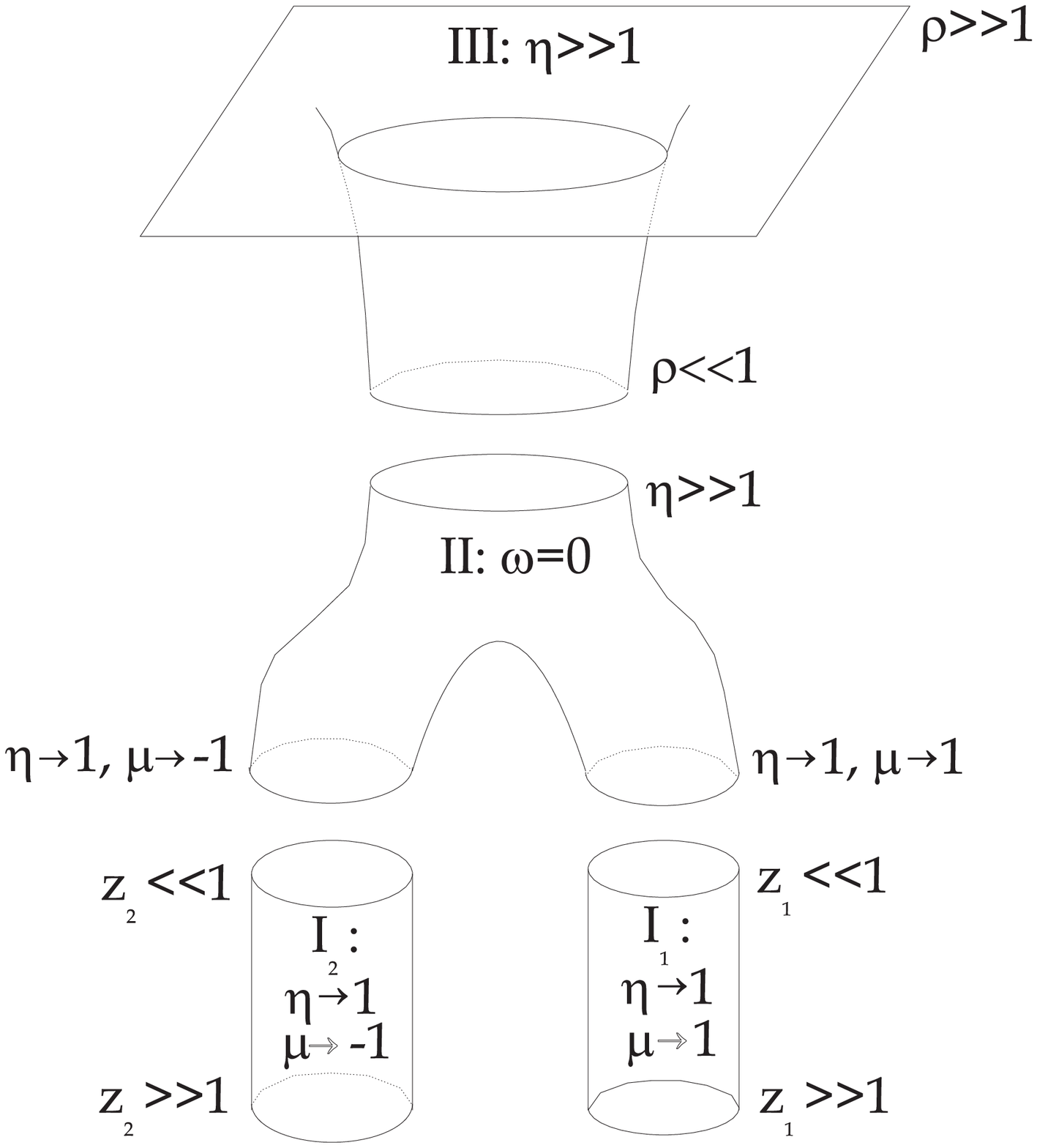,width=3.5in}}
\caption{\small{The different regions used in the solution matching technique.}}
\label{fig2}
\end{figure} 

In this section we shall consider the problem of scattering a minimally coupled scalar
in the double-centered D3-brane background. We assume that the incident wave 
is isotropic with frequency $\o$, i.e. 
\eqn
\phi=e^{-i\o t}\phi(\eta,\mu)\ ,
\label{wave}
\eeqn
with $\phi(\eta,\mu)\sim \phi(\eta)$ for $\eta\gg 1$.
In prolate spherical coordinates the 
Laplace equation reads
\eqn
\left[\grad^2_{(\eta,\mu)}+(\o\D)^2
\left(1+\frac{R_1^{\ 4}}{\D^4(\eta-\mu)^4}+\frac{R_2^{\ 4}}{\D^4(\eta+\mu)^4}\right)\right]
\phi(\eta,\mu)=0\ ,
\label{lapleqn}
\eeqn
where
\eqn
\grad^2_{(\eta,\mu)}=\frac{1}{\eta^2-\mu^2}\left[
\frac{1}{(\eta^2-1)^{\frac{3}{2}}}
\p_{\eta}\left((\eta^2-1)^{\frac{5}{2}}\p_{\eta}\right)
+\frac{1}{(1-\mu^2)^{\frac{3}{2}}}
\p_{\mu}\left((1-\mu^2)^{\frac{5}{2}}\p_{\mu}\right)\right]\ .
\label{lapl}
\eeqn
To solve this equation we use the by now well known 
matching technique. Start by splitting spacetime
in the following four regions: Near-horizon regions ${\bf I_i}$ for $\eta\rightarrow 1$
and $\mu\rightarrow\pm 1$; Intermediate region ${\bf II}$ for $\o=0$; Far region 
${\bf III}$ for $\eta\gg 1$. The different regions are shown in figure 2. The differential
equation (\ref{lapleqn}) becomes in each region
\begin{eqnarray}
{\bf I_i}&:& 
\left[\frac{1}{\z_i^{\ 5}}\p_{\z_i}\left(\z_i^{\ 5}\p_{\z_i}\right)
+(\o\D)^2\left(1+\left(\frac{R_j}{2\D}\right)^4
+\left(\frac{R_i}{\D\z_i}\right)^4\right)\right]\phi(\z_i)=0\ ,
\label{difeqn1}\\
{\bf II}&:&
\grad^2_{(\eta,\mu)}\phi(\eta,\mu)=0\ ,
\label{difeqn2}\\
{\bf III}&:&
\left[\frac{1}{\eta^5}\p_{\eta}\left(\eta^5\p_{\eta}\right)
+(\o\D)^2\left(1+\frac{R_1^{\ 4}+R_2^{\ 4}}{(\D\eta)^4}\right)\right]\phi(\eta)=0\ .
\label{difeqn3}
\end{eqnarray}
Notice that the differential 
equation (\ref{lapleqn}) separates in the four regions. In region ${\bf III}$ we
assumed that the dilaton wave is constant along $S^5$ because we consider a spherically 
symmetric incident wave. Also, in the throat regions ${\bf I_i}$ we assume spherical 
symmetry, i.e. there is no $\varphi_i$ dependence. The reason is that the higher harmonic 
dependence in $\varphi_i$ is related to higher partial waves absorption by each hole 
whose cross-section is smaller by a factor of $(\o R_i)^4$ in comparison to the leading order 
term here calculated. The use of the $\o=0$ equation in region ${\bf II}$ means that we
are scattering the black holes at very low energy. In particular, it means that $\o\D\ll 1$,
i.e. the wave length of the incident wave is much larger than the holes separation distance.

First consider the throat regions ${\bf I_i}$. Defining the coordinate 
$z_i=\frac{\o R_i^{\phantom{i} 2}}{\D\z_i}$ and setting $\phi^{I_i}(z_i)=z_i^{\ 4}\P^{I_i}(z_i)$, 
equation (\ref{difeqn1}) reads
\eqn
\left[\frac{1}{z_i^{\ 5}}\p_{z_i}\left(z_i^{\ 5}\p_{z_i}\right)
+1+\frac{(\o R_i)^4}{z_i^{\ 4}}\left(1+\left(\frac{R_j}{2\D}\right)^4\right)\right]
\P^{I_i}(z_i)=0\ .
\label{difeqn1a}
\eeqn
For $z_i^{\ 2}\gg (\o R_i)^4\left(1+(R_j/2\D)^4\right)$ this can be solved in terms of Bessel 
functions of order 2. The result is
\eqn
\phi^{I_i}(z_i)=z_i^{\ 2}\left(A_iJ_2^{\phantom{1}}(z_i)+iB_iN_2(z_i)\right)\ ,
\label{sol1}
\eeqn
where $A_i$ and $B_i$ are undetermined constants. In the fluxes method for the absorption
cross-section
calculation we require purely infalling flux at the horizons. Setting $A_i=B_i=1$ the
ingoing flux at the horizons is ${\cal F}_{in}(H_i)\cong(\o R_i^{\ 2})^4/\pi$. Notice that
the validity of the approximation in regions ${\bf I_i}$ requires $\z_i\ll 1$ and
$z_i^{\ 2}\gg (\o R_i)^4\left(1+(R_j/2\D)^4\right)$. It is not difficult to see that these
conditions are compatible. To match this solution to the region ${\bf II}$ we require
$z_i\ll 1$ and use the small argument expansion for the Bessel functions. Consistency with
the above conditions requires
\eqn
(\o R_i)^4\left(1+\left(\frac{R_j}{2\D}\right)^4\right)\ll 1
\ \ {\rm and}\ \ 
\frac{\o R_i^{\ 2}}{\D}\ll 1\ .
\label{conds}
\eeqn
As usual we have $(\o R_i)\ll 1$ and the condition $\o R_iR_j/\D\ll 1$ holds for
$\o/m_W\ll 1$. 
\footnote{To be more precise, the condition $\o R_iR_j/\D\ll 1$ is equivalent to
$\o\ll m_g$, where $m_g=m_W/\sqrt{gN}$ is the gravity mass gap. Hence, the domain of
validity of this calculation is restricted to energies $\e\ll m_g\ll m_W$. We shall
comment on a possible field theory interpretation of this fact later.}
In the limit $z_i\ll 1$, (\ref{sol1}) becomes
\eqn
\phi^{I_i}\sim -\frac{4i}{\pi}\ .
\label{asympt1}
\eeqn

In region ${\bf II}$, the solution to the differential equation (\ref{difeqn2})
can be written in terms of the Gegenbauer polynomials  $G^2_l$ as
\eqn
\phi^{II}(\eta,\mu)=\sum_{l=0}^{\infty}C_l\ G^2_l(\eta)\ G^2_l(\mu)\ .
\label{sol2}
\eeqn
However, to match to the asymptotic behavior (\ref{asympt1}), only the constant
term survives
\eqn
\phi^{II}=C=-\frac{4i}{\pi}\ .
\label{sol2a}
\eeqn

Finally, consider the asymptotic flat region ${\bf III}$. Defining the coordinate
$\r=\o\D\eta$ the corresponding differential equation becomes
\eqn
\left[\frac{1}{\r^5}\p_{\r}\left(\r^5\p_{\r}\right)
+1+\o^4\frac{R_1^{\ 4}+R_2^{\ 4}}{\r^4}\right]
\phi^{III}(\r)=0\ .
\label{difeqn3a}
\eeqn
For $\r^2\gg(\o R_i)^4$ the last term is negligible and again this equation can be solved in 
terms of Bessel functions of order 2. The result is
\eqn
\phi^{III}(\r)=\frac{1}{\r^2}\left(DJ_2^{\phantom{1}}(\r)+iEN_2(\r)\right)\ ,
\label{sol3}
\eeqn
where $D$ and $E$ are undetermined constants. 
The validity of the approximation in region ${\bf III}$
requires $\eta\gg 1$ and $\r^2\gg(\o R_i)^4$. These conditions are compatible because
we are assuming that $(\o R_i^{\ 2}/\D)\ll 1$. As usual, in the limit 
$(\o R_i)\ll 1$ we can set $\r\ll 1$ and use the small argument expansion of the Bessel
functions to match the solution to region ${\bf II}$. Notice that the condition $\eta\gg 1$
is compatible with $\r=\o\D\eta\ll 1$ because $\o\D$ is small. In the limit $\r\ll 1$,
(\ref{sol3}) becomes
\eqn
\phi^{III}\sim\frac{D}{8}=-\frac{4i}{\pi}\ ,
\label{asympt3}
\eeqn
where we matched the solution to $\phi^{II}$. Then, the incoming flux at infinity
is seen to be ${\cal F}_{in}(\infty)\cong (16)^2/(\pi^3\o^4)$. 

The probability for absorption by the $i$-th hole can be calculated 
as the ratio of the incoming flux at the $i$-th
horizon to the incoming flux at infinity:
\eqn
{\cal P}_i=\frac{{\cal F}_{in}(H_i)}{{\cal F}_{in}(\infty)}=
\frac{\pi^2}{(16)^2}(\o R_i)^8\ ,
\label{prob}
\eeqn
which is the same result as first obtained by Klebanov in the single-centered 
case \cite{Kleb}. The absorption cross-section is related to this probability 
by $\s_{abs}=(32\pi^2/\o^5){\cal P}$.

\subsection{Logarithmic Corrections to Absorption Cross-Section}

The above calculation provides the leading term in an expansion of the cross-section
in powers of $(\o R_i)$ and $(\o R_i R_j/\D)$.
In this subsection we shall be more careful with the matching conditions across the 
different regions and determine the first corrections to the absorption probability
(\ref{prob}) using the improved matching technique developed in \cite{GHKK}.
At this point, those concerned with the dual field theory interpretation of our
results may wish to skip directly to the final answer (\ref{corrprob}).

First consider the near horizon regions ${\bf I_i}$. Instead of dropping
the last term in equation (\ref{difeqn1a}) we solve this equation by expanding 
$\P^{I_i}(z_i)$ in powers of $(\o R_i)^4\left(1+(R_j/2\D)^4\right)$:
\eqn
\P^{I_i}(z_i)=\P_0^{I_i}(z_i)
+(\o R_i)^4\left(1+\left(\frac{R_j}{2\D}\right)^4\right)\P_1^{I_i}(z_i)
+\cdots\ .
\label{expansion1}
\eeqn
Writing equation (\ref{difeqn1a}) order by order in the expansion parameter we have
\begin{eqnarray}
&&\left[\frac{1}{z_i^{\ 5}}\p_{z_i}\left(z_i^{\ 5}\p_{z_i}\right)+1\right]
\P_0^{I_i}(z_i)=0\ ,
\label{difeqn1.0}\\
&&\left[\frac{1}{z_i^{\ 5}}\p_{z_i}\left(z_i^{\ 5}\p_{z_i}\right)+1\right]
\P_1^{I_i}(z_i)=-\frac{1}{z_i^{\ 4}}\P_0^{I_i}(z_i)\ .
\label{difeqn1.1}
\end{eqnarray}
The solutions to the homogeneous equation are
\eqn
\arr{rcl}
\displaystyle{\frac{J_2(z_i)}{z_i^{\ 2}}}
&\sim&
\displaystyle{\frac{1}{8}\left(1-\frac{z_i^{\ 2}}{12}+{\cal O}(z_i^{\ 4})\right)}\ ,\\
\displaystyle{\frac{N_2(z_i)}{z_i^{\ 2}}}
&\sim&
\displaystyle{-\frac{4}{\pi z_i^{\ 4}}
\left(1+\frac{z_i^{\ 2}}{4}+{\cal O}(z_i^{\ 4}\log{z_i})\right)}\ .
\earr
\label{homosol1}
\eeqn
In the previous calculation we found that 
$\P^{I_i}_0=A_i(J_2+iN_2)/z_i^{\ 2}\equiv A_iH_2^{(1)}/z_i^{\ 2}$.
The solution to the inhomogeneous equation for $\P^{I_i}_1$ can be written in an integral 
form
\eqn
\P^{I_i}_1(z_i)=-\frac{\pi}{2z_i^{\ 2}}\int^{z_i}\frac{d\s}{\s}\P^{I_i}_0(\s)
\left(J_2^{\phantom{1}}(\s)N_2(z_i)-N_2(\s)J_2(z_i)\right)\ ,
\label{ihomosol1}
\eeqn
to which a solution of the homogeneous equation can be added. This ambiguity is resolved
by requiring purely infalling flux at the horizon and by matching to the solution in region
${\bf II}$. The improved solutions in regions ${\bf I_i}$ are given by
\eqn
\arr{rcl}
\displaystyle{\frac{1}{A_i}\phi^{I_i}(z_i)}
&=&
\displaystyle{z_i^{\ 2}H^{(1)}_2(z_i)-
(\o R_i)^4\left(1+\left(\frac{R_j}{2\D}\right)^4\right)\frac{\pi z_i^{\ 2}}{2}}\cdot\\
&&
\displaystyle{\int^{z_i}\frac{d\s}{\s^3}H_2^{(1)}(\s)
\left(J_2^{\phantom{1}}(\s)N_2(z_i)-N_2(\s)J_2(z_i)\right)}\ .
\earr
\label{newsol1}
\eeqn
For small $z_i$ this gives
\eqn
\frac{1}{A_i}\phi^{I_i}(z_i)\sim
-\frac{4i}{\pi}\left[1+\frac{z_i^{\ 2}}{4}
-(\o R_i)^4\left(1+\left(\frac{R_j}{2\D}\right)^4\right)\frac{1}{12z_i^{\ 2}}
\left(1-\frac{z_i^{\ 2}\log{z_i}}{2}\right)+\cdots\right]\ .
\label{newasympt1}
\eeqn

Next we consider the intermediate region ${\bf II}$. The equation for $\phi(\eta,\mu)$ reads
\eqn
\left[\grad^2_{(\eta,\mu)}+(\o\D)^2
\left(1+\frac{R_1^{\ 4}}{\D^4(\eta-\mu)^4}+\frac{R_2^{\ 4}}{\D^4(\eta+\mu)^4}\right)\right]
\phi(\eta,\mu)=0\ ,
\label{againlapleqn}
\eeqn
and in the previous calculation we set $\o=0$. Now we expand the solution in powers of
$(\o\D)^2$:
\eqn
\phi^{II}(\eta,\mu)=\phi_0^{II}(\eta,\mu)+(\o\D)^2\phi_1^{II}(\eta,\mu)
+(\o\D)^4\phi_2^{II}(\eta,\mu)+\cdots\ ,
\label{expansion2}
\eeqn
which gives order by order 
\begin{eqnarray}
&&
\grad^2_{(\eta,\mu)}\phi_0^{II}(\eta,\mu)=0\ ,
\label{difeqn2.0}\\
&&
\grad^2_{(\eta,\mu)}\phi_1^{II}(\eta,\mu)=
-\left(1+\frac{R_1^{\ 4}}{\D^4(\eta-\mu)^4}+\frac{R_2^{\ 4}}{\D^4(\eta+\mu)^4}\right)
\phi_0^{II}(\eta,\mu)\ ,
\label{difeqn2.1}\\
&&
\grad^2_{(\eta,\mu)}\phi_2^{II}(\eta,\mu)=
-\left(1+\frac{R_1^{\ 4}}{\D^4(\eta-\mu)^4}+\frac{R_2^{\ 4}}{\D^4(\eta+\mu)^4}\right)
\phi_1^{II}(\eta,\mu)\ .
\label{difeqn2.2}
\end{eqnarray}
We saw before that $\phi_0^{II}=C$. The other solutions to the homogeneous equation in the
summation (\ref{sol2}) give contributions of order $(\o R_i)^4$ and $(\o R_i^{\ 2}/\D)^4$ to the
matching conditions that we neglect. Hence, the constant $C$ will be chosen to match to
the regions ${\bf I_i}$ and ${\bf III}$ appropriately. Furthermore, we can take the homogeneous 
parts of $\phi_1^{II}$ and $\phi_2^{II}$ to vanish and change the solutions in regions
${\bf I_i}$ and ${\bf III}$ accordingly. This means that we can consider the limits 
$\eta\rightarrow 1$, $\mu\rightarrow\pm 1$ and the limit $\eta\gg 1$ in region ${\bf II}$
and solve there for the inhomogeneous parts of $\phi_1^{II}$ and $\phi_2^{II}$. Start by
considering the limits $\eta\rightarrow 1$, $\mu\rightarrow\pm 1$. Equations
(\ref{difeqn2.1}) and (\ref{difeqn2.2}) become
\begin{eqnarray}
\frac{1}{\z_i^{\ 5}}\p_{\z_i}\left(\z_i^{\ 5}\p_{\z_i}\right)\phi_1^{II_i}
&=&
-\left(1+\left(\frac{R_j}{2\D}\right)^4+\left(\frac{R_i}{\D\z_i}\right)^4\right)C\ ,
\label{difeqn2.1a}\\
\frac{1}{\z_i^{\ 5}}\p_{\z_i}\left(\z_i^{\ 5}\p_{\z_i}\right)\phi_2^{II_i}
&=&
-\left(1+\left(\frac{R_j}{2\D}\right)^4+\left(\frac{R_i}{\D\z_i}\right)^4\right)
\phi_1^{II_i}\ .
\label{difeqn2.2a}
\end{eqnarray}
Solving for these equations we obtain the asymptotic behavior of $\phi^{II}$ as
$\eta\rightarrow 1$, $\mu\rightarrow\pm 1$:
\eqn
\arr{rcl}
\phi^{II_i}&\sim&
\displaystyle{C\left[ 1-(\o\D)^2
\left(1+\left(\frac{R_j}{2\D}\right)^4\right)\frac{\z_i^{\ 2}}{12}
+(\o\D)^2\left(\frac{R_i}{\D}\right)^4\frac{1}{4\z_i^{\ 2}}\right.}\\\\
&&{\displaystyle\left.\ \ \ \ \ \ \ 
-(\o\D)^4\left(1+\left(\frac{R_j}{2\D}\right)^4\right)
\left(\frac{R_i}{\D}\right)^4\frac{1}{24}\log{\z_i}+\cdots\right]}\ .
\earr
\label{newasympt2a}
\eeqn
Similarly, in the limit $\eta\gg 1$ equations (\ref{difeqn2.1}) and (\ref{difeqn2.2}) 
become
\begin{eqnarray}
\frac{1}{\eta^5}\p_{\eta}\left(\eta^5\p_{\eta}\right)\phi_1^{II_{\infty}}
&=&
-\left(1+\frac{R_1^{\ 4}+R_2^{\ 4}}{(\D\eta)^4}\right)C\ ,
\label{difeqn2.1b}\\
\frac{1}{\eta^5}\p_{\eta}\left(\eta^5\p_{\eta}\right)\phi_2^{II_{\infty}}
&=&
-\left(1+\frac{R_1^{\ 4}+R_2^{\ 4}}{(\D\eta)^4}\right)\phi_1^{II_{\infty}}\ ,
\label{difeqn2.2b}
\end{eqnarray}
which gives the following asymptotic behavior of $\phi^{II}$ for $\eta\gg 1$:
\eqn
\arr{rcl}
\phi^{II_{\infty}}
&\sim& 
\displaystyle{C\left[1-(\o\D)^2\frac{\eta^2}{12}
+(\o\D)^2\frac{R_1^{\ 4}+R_2^{\ 4}}{\D^4}\frac{1}{4\eta^2}\right.}\\\\
&&\displaystyle{\ \ \ \ \ \ \ \left.
-(\o\D)^4\frac{R_1^{\ 4}+R_2^{\ 4}}{\D^4}\frac{1}{24}\log{\eta}+\cdots\right]}\ .
\label{newasympt2b}
\earr
\eeqn

Finally, consider the far region {\bf III}. Instead of dropping the last term
in equation (\ref{difeqn3a}), we solve this equation by expanding $\phi^{III}(\r)$
in powers of $\o^4(R_1^{\ 4}+R_2^{\ 4})$:
\eqn
\phi^{III}(\r)=\phi_0^{III}(\r)
+\o^4\left(R_1^{\ 4}+R_2^{\ 4}\right)\phi_1^{III}(\r)+\cdots\ .
\label{expansion3}
\eeqn
Then, the analysis is entirely similar to the one performed for the near horizon
regions ${\bf I_i}$, therefore we present just the final result
\eqn
\frac{1}{D}\phi^{III}(\r)=
\frac{J_2(\r)}{\r^2}
-\o^4\left( R_1^{\ 4}+R_2^{\ 4} \right)\frac{\pi}{2\r^2}
\int^{\r}\frac{d\s}{\s^3}J_2(\s)\left(J_2^{\phantom{1}}(\s)N_2(\r)-N_2(\s)J_2(\r)\right)\ .
\label{newsol3}
\eeqn
For small $\r$ this gives
\eqn
\frac{1}{D}\phi^{III}(\r)\sim
\frac{1}{8}\left( 1-\frac{\r^2}{12}\right)
+\frac{\o^4\left( R_1^{\ 4}+R_2^{\ 4} \right)}{32\r^2}
\left( 1-\frac{1}{6}\r^2\log{\r}\right)+\cdots\ .
\label{newasympt3}
\eeqn

Now we are in position to determine the improved matching conditions across the different
regions. To match the expansions (\ref{newasympt1}) and (\ref{newasympt2a}) for $\phi^{I_i}$
and $\phi^{II_i}$ we define the variable $u_i$ by 
\eqn
u_i=\frac{R_i}{\D}\z_i\ ,\ \ \ z_i=(\o R_i)\left(\frac{R_i}{\D}\right)^2\frac{1}{u_i}\ ,
\label{matchvar1}
\eeqn
where we assumed that $R_i\gg \D$. Notice that this condition is necessary in order
to keep $u_i$ finite of order unit in the domain of validity of the expansions for 
$\phi^{I_i}$ and $\phi^{II_i}$. Replacing (\ref{matchvar1}) in these expansions we have
\eqn
\arr{rcl}
\displaystyle{\frac{1}{A_i}\phi^{I_i}}&\sim&
\displaystyle{-\frac{4i}{\pi}\left[1+(\o R_i)^2
\left(\frac{R_i}{\D}\right)^4\frac{1}{4u_i^{\ 2}}
-(\o R_i)^2\left(1+\left(\frac{R_j}{2\D}\right)^4\right)
\left(\frac{\D}{R_i}\right)^4\frac{u_i^{\ 2}}{12}\right.}\\\\
&&\displaystyle{\left.
+(\o R_i)^4\left(1+\left(\frac{R_j}{2\D}\right)^4\right)\frac{1}{24}
\left(\log{\left(\frac{\o R_i^{\ 2}}{\D}\right)}-\log{\left(\frac{\D u_i}{R_i}\right)}\right)
+\cdots\right]}\ ,\\\\
\displaystyle{\frac{1}{C}\phi^{II_i}}&\sim&
\displaystyle{1+(\o R_i)^2
\left(\frac{R_i}{\D}\right)^4\frac{1}{4u_i^{\ 2}}
-(\o R_i)^2\left(1+\left(\frac{R_j}{2\D}\right)^4\right)
\left(\frac{\D}{R_i}\right)^4\frac{u_i^{\ 2}}{12}}\\\\
&&\displaystyle{
-(\o R_i)^4\left(1+\left(\frac{R_j}{2\D}\right)^4\right)\frac{1}{24}
\log{\left(\frac{\D u_i}{R_i}\right)}+\cdots}\ .
\earr
\label{match1}
\eeqn
By adjusting the constants $A_i$, we can match the logarithmic terms in the expansions:
\eqn
A_i=1-\frac{(\o R_i)^4}{24}\left(1+\left(\frac{R_j}{2\D}\right)^4\right)
\log{\left(\frac{\o R_i^{\ 2}}{\D}\right)}\ ,\ \ \ C=-\frac{4i}{\pi}\ .
\label{matchconst1}
\eeqn
The next corrections will be of order $(\o R_i)^4$ and $(\o R_i R_j/\D)^4$, but we 
neglect these.
In the case $R_i\ll \D$, we define instead $u_i=\frac{\D}{R_i}\z_i$ which can be kept
finite in the matching region. Of course the constants $A_i$ and $C$ will still be given 
by (\ref{matchconst1}). 

Similarly, to match the expansions (\ref{newasympt2b}) and 
(\ref{newasympt3}) for $\phi^{II_{\infty}}$
and $\phi^{III}$, we define the variable $u$ by 
\eqn
u=\frac{\D}{R_1+R_2}\eta\ ,\ \ \ \r=\o\left( R_1+R_2\right)u\ ,
\label{matchvar2}
\eeqn
where we assumed that $R_i\gg \D$. Across the matching region, $u$ can be kept fixed
of order unit and the expansions $\phi^{II_{\infty}}$ and $\phi^{III}$ read
\eqn
\arr{rcl}
\displaystyle{\frac{1}{C}\phi^{II_{\infty}}}&\sim&
\displaystyle{1-\o^2\left(R_1+R_2\right)^2\frac{u^2}{12}
+\o^2\frac{R_1^{\ 4}+R_2^{\ 4}}{\left(R_1+R_2\right)^2}\frac{1}{4u^2}}\\\\
&&
\displaystyle{-\o^4\left( R_1^{\ 4}+R_2^{\ 4} \right)\frac{1}{24}
\log{\left(\frac{R_1+R_2}{\D}u\right)}+\cdots}\ ,\\\\
\displaystyle{\frac{1}{D}\phi^{III}}&\sim&
\displaystyle{\frac{1}{8}
\left[1-\o^2\left(R_1+R_2\right)^2\frac{u^2}{12}
+\o^2\frac{R_1^{\ 4}+R_2^{\ 4}}{\left(R_1+R_2\right)^2}\frac{1}{4u^2}\right.}\\\\
&&
\displaystyle{\left.\phantom{\frac{R_1^1}{R_1^1}}
-\o^4\left( R_1^{\ 4}+R_2^{\ 4} \right)\frac{1}{24}
\log{\left(\o\left(R_1^{\phantom{1}}+R_2\right)u\right)\cdots}\right]}\ .
\earr
\label{match2}
\eeqn
By adjusting the constant $D$ we can match the logarithmic terms in the expansions:
\eqn
D=-\frac{32i}{\pi}
\left[1+\frac{\o^4}{24}\left(R_1^{\ 4}+R_2^{\ 4}\right)\log{\left(\o\D\right)}\right]\ .
\label{matchconst2}
\eeqn
In the case $R_i\ll\D$, we define instead $u=\frac{R_1+R_2}{\D}\eta$ which can be kept
finite across the matching region.

With the improved matching conditions we can calculate the fluxes at the horizons
and at infinity. Then, the corrected absorption 
probability by the $i$-th hole is seen to be
\eqn
{\cal P}_i=
\frac{\pi^2(\o R_i)^8}{(16)^2}\left[ 
1-\frac{(\o R_i)^4}{12}\left(1+\left(\frac{R_j}{2\D}\right)^4\right)
\log{\left(\frac{\o R_i^{\ 2}}{\D}\right)}
-\frac{\o^4\left(R_1^{\ 4}+R_2^{\ 4}\right)}{12}\log{\left(\o\D\right)}
\right]\ ,
\label{corrprob}
\eeqn
which holds either for $R_i\gg\D$ or $R_i\ll\D$.
This is our main result. In the reminder of this paper we interpret these corrections
to the absorption probability from a field theory point of view.

As an aside, notice that this calculation can be easily extended to the multi-centered 
D3-brane geometry case. The dual field theory is the large $N_i$ SYM theory at strong 
coupling with gauge group broken to $S(\prod_i U(N_i))$. Then, the absorption 
probability by the $i$-th hole is
\eqn
{\cal P}_i=
\frac{\pi^2(\o R_i)^8}{(16)^2}\left[ 
1-\frac{(\o R_i)^4}{12}\left(1+\sum_{j\ne i}\left(\frac{R_j}{\D_{ij}}\right)^4\right)
\log{\left(\frac{\o R_i^{\ 2}}{\D}\right)}
-\frac{\o^4\sum_iR_i^{\ 4}}{12}\log{\left(\o\D\right)}
\right]\ ,
\label{multi}
\eeqn
where $\D_{ij}$ is the separation between the $i$-th and $j$-th D3-branes and $\D$ is a
typical brane separation. The field theory analysis of this case is entirely similar 
to the double-centered case presented in the next section. 

\section{Field Theory Analysis}
\news

We are interested in maximally supersymmetric RG flows for ${\cal N}=4$ SYM theory
that arise from breaking
the conformal invariance. In the infrared it has been conjectured that these
flows can be viewed as a deformation of the ${\cal N}=4$ superconformal fixed points
by some irrelevant operators \cite{GHKK,GubserHash,Intr}. In particular, the gauge 
coupling beta function vanishes and the dimension of operators in short supersymmetry
representations are constant along such flows. 

First consider the D3-brane background with $SO(6)\cong SU(4)$ invariant harmonic
function
\eqn
H=h+\left(\frac{R}{r}\right)^4\ .
\label{harm1}
\eeqn
Interpreting $r$ as the RG energy scale the dual field theory flows in the IR 
($r\rightarrow 0$) to the ${\cal N}=4$ superconformal fixed point. The corresponding
deformation, which is controlled by the parameter $h$, can be seen to arise from the
irrelevant Dirac-Born-Infeld (DBI) corrections to the SYM theory \cite{GHKK}. 
Alternatively, on the basis of $PSU(2,2|4)$ representation theory the 
Lagrangian of the dual field theory was conjectured to be \cite{GubserHash,Intr}
\eqn
{\cal L}={\cal L}_0-\frac{h}{8T_3}{\cal O}_8\ ,
\label{Lagr}
\eeqn
where
\eqn
{\cal L}_0=
-\frac{1}{4}{\rm Tr}\left(F_{AB}F^{AB} \right)\equiv
-\frac{1}{4}{\cal O}_4\ ,
\label{SYM}
\eeqn
is the usual bosonic SYM Lagrangian and ${\cal O}_8$ is a dimension 8 operator
preserving 16 supersymmetries and the $SU(4)_R$ symmetry. We use the ten-dimensional
indices $A,B$ so that $F_{AB}$ is short for $F_{\m\n}=D_{\m}A_{\n}-D_{\n}A_{\m}$, 
$F_{\m m}=D_{\m}\phi_m$ with $D_{\m}\equiv \p_{\m}+ig_{YM}[A_{\m},\ ]$ and
$F_{mn}=ig_{YM}[\phi_m,\phi_n]$. Notice that we removed the gauge coupling from the
front of the action by rescalling the fields according to 
$(A_{\m},\phi^m)\rightarrow g_{YM}(A_{\m},\phi^m)$.
The D3-brane tension $T_3$ is given by $T_3=((2\pi)^3g_s\a'^2)^{-1}$.

In the case where there are two stacks of D3-branes separated by a distance $2\D$,
the harmonic function (\ref{harm1}) becomes
\eqn
H=h+\frac{R_1^{\ 4}}{|\vec{r}-\vec{\D}|^4}
+\frac{R_2^{\ 4}}{|\vec{r}+\vec{\D}|^4}\ .
\label{harm2}
\eeqn
The corresponding dual field theory is still given by (\ref{Lagr}) but the $SU(N)$ gauge
symmetry is broken to $S(U(N_1)\times U(N_2))$ by the non-zero expectation values
for the scalar fields as described 
in the Introduction. For example, this means that there will be a mass term in the 
Lagrangian ${\cal L}_0$ for the W particles. 

Our main goal is to identify the origin and corresponding
physical processes associated with the logarithmic corrections to the cross-section 
calculated in the previous section. For worldvolume on-shell processes involving 
the coupling of the dilaton field to the brane, the important terms in the Lagrangian 
(\ref{Lagr}) that arise from the DBI action are
\eqn
\arr{rcl}
{\cal O}_4 &=& {\rm Tr}\left(F^{\phantom{1}}_{\m\n}F^{\m\n}\right)\ ,\\
{\cal O}_8 &=& {\rm Tr}\left(F_{\m\n}F^{\n\r}F_{\r\s}F^{\s\m}
-\frac{1}{4}\left(F^{\phantom{1}}_{\m\n}F^{\m\n}\right)^2\right)\ .
\earr
\label{oper}
\eeqn
Also, the coupling of the dilaton field to the brane world-volume is seen to be
\eqn
S_{int}=\int d^4x\ \phi\ {\cal O}_{\phi}
=\int d^4x\ \phi\ \frac{1}{4}\left[{\cal O}_4+\frac{h}{T_3}{\cal O}_8+\cdots\right]\ ,
\label{inter}
\eeqn
where ${\cal O}_4$ and ${\cal O}_8$ are defined in (\ref{oper}). 
Notice that a more careful analysis to account for numerical factors 
would require the use of the symmetrized trace as 
proposed by Tseytlin \cite{Tsey}.

To calculated the cross-section 
one starts by calculating the two-point function of the operator 
${\cal O}_{\phi}$
\eqn
\Pi(x)=\left\langle {\cal O}_{\phi}^{\phantom{1}}(x){\cal O}_{\phi}(0)\right\rangle_h\ ,
\label{2pf}
\eeqn
where the subscript $h$ reminds us that we are working with the deformed Lagrangian 
(\ref{Lagr}). Then, the absorption cross-section is given by \cite{GubserKleb}
\eqn
\s=\frac{2\kappa^2}{2i\o}\left.{\rm Disc}^{\phantom{1}}
\Pi(s)\right|_{{p^0 = \omega  \atop \vec{p} =\vec{0}}}\ ,
\label{csection}
\eeqn
where $s=-p^2$ and $\Pi(s)$ is the Fourier transform
of $\Pi(x)$. The momentum space two-point function $\Pi(s)$ is analytic in the complex
$s$-plane except possibly for poles or cuts on the real axis. The discontinuity across
the cuts is given by ${\rm Disc}\ \Pi(s)=\Pi(s+i\e)-\Pi(s-i\e)$.  

Now that we have reviewed the connection between the classical cross-section and the dual field 
theory, we interpret the logarithmic corrections to the cross-section found earlier from
the field theory point of view. We shall consider the two cases $R_i\gg \D$ and $R_i\ll \D$
discussed in Section 2 separately.

\subsection{Small Brane Separation ($R_i\gg \D$)}

First, consider the case where $\D/\a'$ is kept fixed in the $\a'\rightarrow 0$ 
decoupling limit. The gravitational dual geometry is sketched in figure 1(c)
of Section 2. In this case the absorption probability becomes 
\eqn
{\cal P}_i=\frac{\pi^2(\o R_i)^8}{(16)^2}
\left[ 1-\frac{(\o R_i)^4}{12}\left(\frac{R_j}{2\D}\right)^4
\log{\left(\frac{\o R_i^{\ 2}}{\D}\right)}\right]\ .
\label{csection2}
\eeqn
The tree-like throat geometry is dual to the ${\cal N}=4$ SYM theory away from the
origin of the moduli space of vacua, i.e. there is a VEV for the Higgs field that
breaks conformal invariance and the gauge symmetry to $S(U(N_1)\times U(N_2))$. The 
corresponding harmonic function in the gravity solution is given by (\ref{harm2}) with
$h=0$. The theory flows from the UV conformal point with gauge group $SU(N_1+N_2)$
to the IR conformal fixed points with gauge group $SU(N_i)$. 

In the gravity calculation we assumed $\o\ll m_W$, which means that we are studying 
deformations of the IR fixed points. In the case of the $i$-th throat, we
are studying the RG flow near the $SU(N_i)$ IR fixed point with conformal invariance
broken due to a deformation by an irrelevant operator of dimension 8. 
In fact, down the throat the harmonic function becomes effectively \cite{Intr}
\eqn
H=\left(\frac{R_j}{2\D}\right)^4+\left(\frac{R_i}{r}\right)^4\ ,
\label{harm3}
\eeqn
where $r\equiv\D\z_i$ is the radial coordinate centered at the $i$-th stack of branes.
We conclude that in this limit the theory is described by the Lagrangian 
(\ref{Lagr}) with $h=(R_j/2\D)^4$. The deformation in this Lagrangian is the 
result of integrating out the massive W particles, defining an effective Lagrangian
near the IR fixed point. This is analogous to the situation
where a brane is placed in the near-horizon geometry of a large number of branes. 
There, the correct action describing the probe dynamics is the DBI action, which
effectively describes the effect of integrating out all the massive strings
stretching between the probe and the branes $[33-36]$. 

To make the idea of the previous paragraph more precise consider the 
$SU(N_1+N_2)$ SYM theory with bosonic Lagrangian given by (\ref{SYM}).
The VEV for the scalars  
$g_{YM}\langle\vec{\phi}\rangle=\frac{\vec{\D}}{2\pi\a'}{\rm Diag}(I_1,-I_2)$ breaks conformal
invariance and the gauge symmetry to $S(U(N_1)\times U(N_2))$. For energies $\e\ll m_W$,
we can integrate out the W particles and obtain an effective action for the $SU(N_i)$
light degrees of freedom. This is similar to the case $N_1=N$ and $N_2=1$ that 
has been extensively studied in the literature $[33-36]$. 
We refer the reader to these references for the details. Since we are interested 
in the large $N$ limit only the planar diagrams contribute to the effective
potential. The first non-vanishing one-loop diagram
involves 4 $SU(N_i)$ coloured external legs. In our conventions, the resulting
contribution to the Lagrangian is \cite{Mald2}
\eqn
{\cal L}_1=-\frac{\pi^2g^4_{YM}}{(2\D/\a')^4}N_j
{\rm Tr}\left(F_{AB}F^{BC}F_{CD}F^{DA}-\frac{1}{4}\left(F_{AB}F^{AB}\right)^2\right)\ ,
\label{1loop}
\eeqn
where $g^2_{YM}=2\pi g_s$ and the factor $N_j$ is because we have a 
$SU(N_j)$ colour index around the loop to sum over. Notice that the fields 
in this expression take values in the $SU(N_i)$ Lie algebra. We see that for 
$h=(R_j/2\D)^4=(2\a'^2g_{YM}^2N_j)/(2\D)^4$ the one-loop result (\ref{1loop}) gives 
exactly the same answer as (\ref{Lagr}) if we take the irrelevant
operator ${\cal O}_8$ to arise from the DBI corrections. 
Furthermore, the $F^4$ terms are exact because they are not renormalized by higher 
loop diagrams \cite{DineSeiberg}. To summarize, we start with 
pure SYM theory and break conformal invariance by going to the Coulomb branch of the theory. 
By integrating out the massive W particles we derived an effective Lagrangian near the
IR fixed points that has the form conjectured in \cite{GHKK,GubserHash,Intr}.

Following the analysis of \cite{GHKK} the two-point function for the operator 
${\cal O}_{\phi}$ dual to the dilaton field is
\eqn
\arr{rcl}
\displaystyle{
\left\langle {\cal O}_{\phi}^{\phantom{1}}(x){\cal O}_{\phi}(0)
\right\rangle_{h=\left(R_j/2\D\right)^4}}
&=&
\displaystyle{\int{\cal D}A_{\mu}
e^{-\int d^4z\left[\frac{1}{4}{\cal O}_4+\frac{h}{8T_3}{\cal O}_8\right]}
{\cal O}_{\phi}(x){\cal O}_{\phi}(0)}\\
&=&
\displaystyle{\left\langle {\cal O}_{\phi}(x){\cal O}_{\phi}(0) 
\left( 1-\left(\frac{R_j}{2\D}\right)^4\frac{1}{8T_3}
\int d^4z{\cal O}_8(z) \right)\right\rangle_{h=0}}\\
&=&
\displaystyle{\frac{1}{2^4}
\left\langle{\cal O}_4^{\phantom{1}}(x){\cal O}_4(0)\right\rangle_{h=0}}\\
&&
\displaystyle{
-\left(\frac{R_j}{2\D}\right)^4\frac{1}{2^7T_3}\int d^4z
\left\langle{\cal O}_4^{\phantom{1}}(x){\cal O}_8(z){\cal O}_4(0)\right\rangle_{h=0}}\ ,
\earr
\label{2pfb}
\eeqn
where we are just keeping the term that will give the $log$ correction calculated in the
gravity approach. Notice that the two-point function for the operator 
${\cal O}_{\phi}$ is written in terms of two- and three-point functions of chiral
primary operators at the IR $SU(N_i)$ conformal fixed point. Since these are expected
to be protected [32,38-43] the perturbative field theory and 
strongly coupled gravity approaches should give the same result.
Now the analysis is entirely similar to the one presented in 
\cite{GHKK}. We refer the reader to that paper for the details. 
The first term in the last equality 
gives the leading term in the scattering cross-section, while 
the second term gives the logarithmic correction. The latter is associated with 
a process where a virtual pair of gauge particles interacts with the incident wave to 
create a pair of gauge particles. The strength of such interaction is determined
by the effective action that arised from integrating out the W particles. To match
the logarithmic term in the field theory calculation we should set the ultraviolet 
cut off to $\Lambda=\D/R_i^{\ 2}$. This is related to the natural scale in the field theory
side $m_W$ by $\Lambda\sim m_W/\sqrt{gN_i}$. A better understanding of the ultraviolet 
cut off requires extending the cross-section calculation to the next order \cite{GHKK}.
Notice, however, that the scale $\Lambda=\D/R_i^{\ 2}$ is the gravity mass gap and may be
related to the existence of colour singlet condensates of W particles at strong 
coupling \cite{GiddRoss,Gubser}. 

\subsection{Large Brane Separation ($R_i\ll \D$)}

As explained in Section 2 the two throats can be separated in the decoupling limit 
as long as $\D/\a'\rightarrow \infty$ (recall figures 1(c) and 1(d)). 
Here we consider the effects 
of connecting both throats using the full D3-brane geometry with harmonic function 
given by (\ref{harm2}) with $h=1$. In this case the absorption probability becomes
\eqn
{\cal P}_i=\frac{\pi^2(\o R_i)^8}{(16)^2}
\left[ 1-\frac{(\o R_i)^4}{6}\log{\left(\o R_i\right)}
-\frac{(\o R_j)^4}{12}\log{\left(\o\D\right)}\right]\ .
\label{csection1}
\eeqn
In the dual field theory we start with the action (\ref{Lagr}) with gauge group 
$SU(N_1+N_2)$ and $h=1$. Additionally, the VEV for the Higgs field breaks the gauge 
symmetry to $S(U(N_1)\times U(N_2))$. In other words, we consider the Coulomb branch of the
deformed theory. This is very similar to the 
single-centered case studied in \cite{GHKK} where conformal invariance was 
broken by setting $h=1$ while remaining at the origin of the moduli space of vacua. 
Intriligator argued that in the UV this theory
is dual to the asymptotically flat space. In the IR we end up with the $SU(N_i)$
IR conformal fixed points corresponding to each throat. 

\begin{figure}
\centerline{\psfig{figure=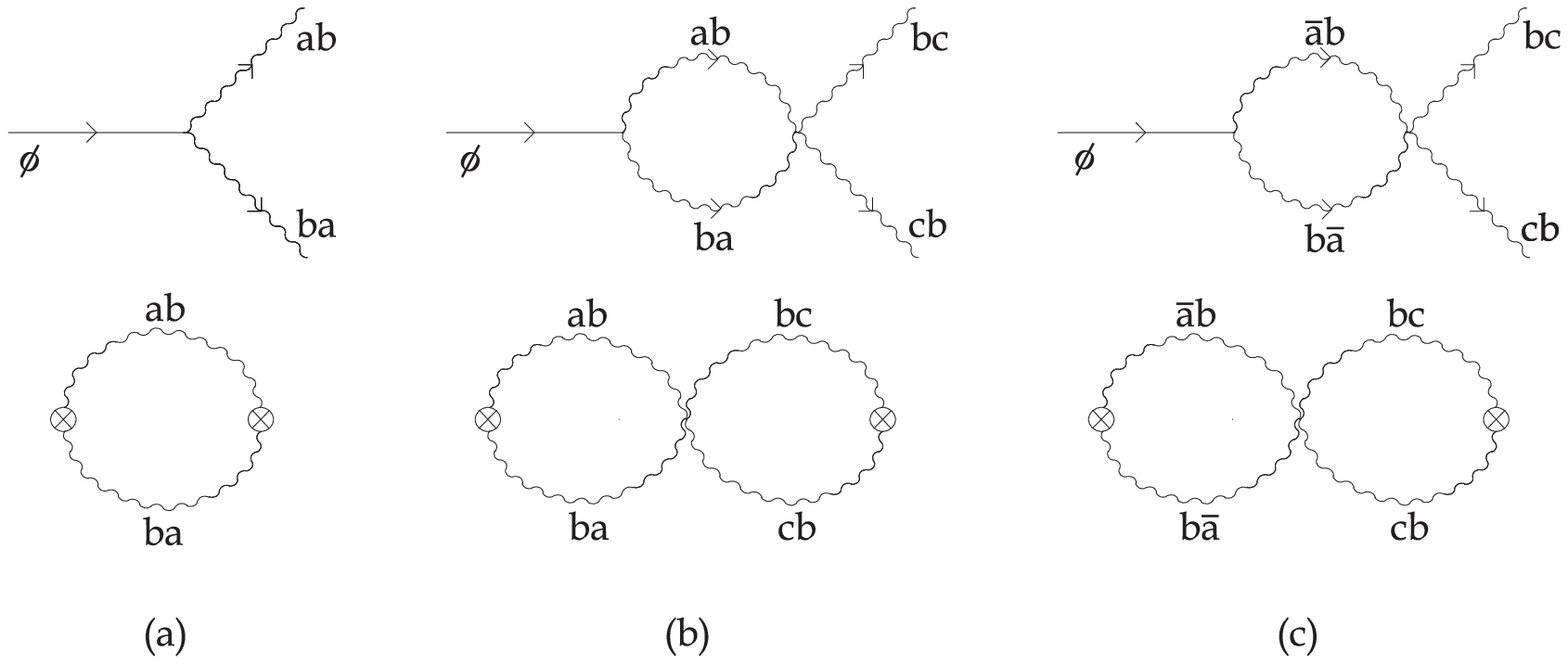,width=6in}}
\caption{\small{Feynman diagrams for the absorption cross-section by the first black hole. 
After summing over the YM indices we see that the leading order term arising
from the diagram (a) is proportional to $(N_1)^2$ and that the logarithmic corrections that 
arise from the closed loops in the top diagrams of (b) and (c) are proportional to
$(N_1)^3$ and $(N_1)^2N_2$, respectively. This is in agreement with the gravity 
calculation.}}
\label{fig3}
\end{figure}

It is convenient to write the operators ${\cal O}_4$ and ${\cal O}_8$ in (\ref{oper})
using the following notation
\eqn
\arr{rcl}
{\cal O}_4\equiv {\cal O}_4^{\g_1}+{\cal O}_4^W+{\cal O}_4^{\g_2}
&\sim&
\displaystyle{\sum_{a,b}F^{ab}F^{ba}+2\sum_{a,\bar{b}}F^{a\bar{b}}F^{\bar{b}a}
+\sum_{\bar{a},\bar{b}}F^{\bar{a}\bar{b}}F^{\bar{b}\bar{a}}}\ ,\\
{\cal O}_8\equiv {\cal O}_8^{\g_1}+{\cal O}_8^{\g_1W}+{\cal O}_8^{W}+\cdots
&\sim&
\displaystyle{\sum_{a,b,c,d}F^{ab}F^{bc}F^{cd}F^{da}
+4\sum_{\bar{a},b,c,d}F^{\bar{a}b}F^{bc}F^{cd}F^{d\bar{a}}}\\
&&
\displaystyle{+2\sum_{\bar{a},b,\bar{c},d}
F^{\bar{a}b}F^{b\bar{c}}F^{\bar{c}d}F^{d\bar{a}}+\cdots}\ ,
\earr
\label{YMoper}
\eeqn
where the indices $a,b,...$ run over the $SU(N_1)$ gauge group and the indices 
$\bar{a},\bar{b},...$ over the $SU(N_2)$ gauge group. In the expression for ${\cal O}_8$
we wrote just the terms that are important for the absorption by the first hole.
Also, we dropped a four point vertex with one $SU(N_1)$ gauge particle,
one $SU(N_2)$ gauge particle and two W bosons 
($\sim F^{\bar{a}\bar{b}}F^{\bar{b}c}F^{cd}F^{d\bar{a}}$) because it gives a
negligible correction for large $N$.
Similarly to the previous section, the two-point function for the operator 
${\cal O}_{\phi}$ is
\eqn
\arr{rcl}
\left\langle {\cal O}_{\phi}^{\phantom{1}}(x){\cal O}_{\phi}(0)\right\rangle_{h=1}
&=&
\displaystyle{\int{\cal D}A_{\mu}
e^{-\int d^4z\left[\frac{1}{4}{\cal O}_4+\frac{1}{8T_3}{\cal O}_8\right]}
{\cal O}_{\phi}(x){\cal O}_{\phi}(0)}\\
&=&
\displaystyle{\left\langle {\cal O}_{\phi}(x){\cal O}_{\phi}(0) 
\left( 1-\frac{1}{8T_3}\int d^4z{\cal O}_8(z) \right)\right\rangle_{h=0}}\\
&=&
\displaystyle{\sum_{i=1}^2\left[
\frac{1}{2^4}\left\langle{\cal O}_{4_{\phantom{1}}}^{\g_i}(x)
{\cal O}_4^{\g_i}(0)\right\rangle_{h=0}
-\frac{1}{2^7T_3}\int d^4z
\left\langle{\cal O}_{4_{\phantom{1}}}^{\g_i}(x){\cal O}_8^{\g_i}(z)
{\cal O}_4^{\g_i}(0)\right\rangle_{h=0}
\right.}\\
&&
\displaystyle{\ \ \ \ \ \ \left.-\frac{1}{2^6T_3}\int d^4z
\left\langle{\cal O}_4^{W}(x){\cal O}_8^{\g_iW}(z){\cal O}_4^{\g_i}(0)\right\rangle_{h=0}
\right]}\\
&&
\displaystyle{
+\frac{1}{2^4}\left\langle{\cal O}_4^{W}(x){\cal O}_4^{W}(0)\right\rangle_{h=0}
-\frac{1}{2^7T_3}\int d^4z
\left\langle{\cal O}_4^{W}(x){\cal O}_8^{W}(z){\cal O}_4^{W}(0)\right\rangle_{h=0}}\ ,
\earr
\label{2pfa}
\eeqn
where we are just keeping the leading terms.
First consider the terms inside the squared brackets in the last equality. 
The first term is the two-point function for the $SU(N_i)$ theory at the infrared conformal 
fixed point and it gives rise to the leading term in the absorption probability 
(\ref{csection1}). This term is 
associated with a process where the incident wave interacts with the brane to create
a pair of gauge particles  (see figure 3(a)) \cite{Kleb}. The first logarithmic correction in the
absorption probability (\ref{csection1}) arises from the second term in the squared brackets. This
correction was analyzed in \cite{GHKK} and it is associated with a process where a virtual pair
of gauge particles interacts with the incident dilaton wave to create a pair of gauge
particles (see figure 3(b)). Finally, the other logarithmic correction in the cross-section
arises from the third term in the squared brackets. This corresponds to a process where 
a virtual pair of W bosons interacts with the incident dilaton wave to create a pair 
of gauge particles (see figure 3(c)). The last two field theory processes have the correct
energy and $N_i$ dependence to account for the logarithmic corrections to the cross section.
However, how to match precisely the numerical factors remains an open question. 
As observed in the previous subsection, an appropriate
understanding of the UV cut off would require extending the analysis to the next order.

The cross-section for all the above processes is calculated
from the discontinuity of the momentum space two-point function across the positive 
real axis in the $s$-plane. In fact, the correction
$\int d^4z\langle{\cal O}_4^{W}(x){\cal O}_8^{\g_iW}(z){\cal O}_4^{\g_i}(0)\rangle_{h=0}$
is also associated to the creation of a pair of W bosons via a virtual pair of
gauge particles. In this case the momentum space two-point function has an 
extra discontinuity in the $s$-plane over the real axis for $s\ge m_W$. Since we work
at energies $\o\ll m_W$, well below the threshold where the channel for W boson pair
production opens, this process is not seen in the gravity calculation here employed. 
Similarly, the Fourier transform of the terms outside the squared
brackets in (\ref{2pfa}) will have a cut for $s\ge m_W$ and are associated with the 
creation of W boson pairs.  

\section{Concluding Remarks}

The main goal of this work was to investigate the Coulomb branch of ${\cal N}=4$ SYM 
theory using the AdS/CFT correspondence. This was done by studying the scattering
of a minimally coupled scalar field by two parallel stacks of D3-branes. 
Technical reasons forced us
to consider processes at energy scales much smaller than the threshold
for W boson pair production. Despite this fact we were still able to find new
phenomena in this phase of the theory. Firstly, for small brane separation, 
we considered the field theory dual to the tree-like throat geometry. This is 
a point on the Coulomb branch of ${\cal N}=4$ SYM theory. Near the IR fixed 
points the theory can be described by integrating out the massive W particles.
As a result, the effective Lagrangian leads to a correction to the 
cross-section that is associated with the creation of gauge particles through 
the production of virtual pairs of gauge particles. This result is perfectly
consistent with the classical gravity calculation. Secondly,
for large brane separation, we considered the
theory that results from connecting both throats with asymptotic flat space. 
This is associated with the DBI corrections to the SYM theory. In this case
we found corrections to the cross-section that are associated with the creation
of gauge particle pairs through the production of virtual pairs of gauge particles 
and of W bosons. For both small or large brane separation the W particles that arise 
from the $SU(N)\rightarrow S(U(N_1)\times SU(N_2))$ symmetry breaking process are seen 
to play an important role in the gravity/gauge theory correspondence. While these particles
are essential to identify the appropriate field theory processes associated with the
logarithmic corrections to the classical cross section, the problem of matching exactly
the numerical factors remains unsolved.

It would be very interesting to extend our results to higher energies. In particular,
one would like to see the channel for W bosons pair production to open up in the
gravitational side of the correspondence. However, the scale in the gravity side is 
given by the mass gap energy $m_g=m_W/\sqrt{gN}$. As noticed earlier, this fact is 
expected to be related to the existence of colour singlets of BPS stretched strings
with large binding energy. Hence, by studying the classical absorption for 
energies $\e\sim m_g$ we could obtain some information about these bound states.
Maybe our best hope would be to study numerically the solutions of Laplace 
equation in the double-centered D3-brane geometry. As far as the W particles are
concerned, it looks that in the gravity limit these particles are not seen directly
and that we may need the full string theory on $AdS_5\times S^5$ to recover them. 

An obvious extension of this work is to consider the scalar absorption by the 
double centered M-brane geometries. Unfortunately, in this case we do not have the input
from D-brane and SYM physics to fully understand the dual field theory picture. 
Another interesting venue of research would be to study the four- and five-dimensional 
black hole cases. These are related to the $AdS_2/CFT_1$ and $AdS_3/CFT_2$ dualities,
respectively.

\section*{Acknowledgments}

I would like to thank Lori Paniak and Igor Klebanov for many discussions and 
for reading a draft of the paper. I am specially thankful to Igor Klebanov
for making many important suggestions. This work was supported by FCT (Portugal) 
under programme PRAXIS XXI and by the NSF grant PHY-9802484.


\newpage

\begin{thebibliography}{40}

\bibitem{Mald}J.M. Maldacena, {\em The Large N Limit of Superconformal Field 
Theories and Supergravity}, Adv. Theor. Math. Phys. {\bf 2} (1998) 231, hep-th/9711200.

\bibitem{GKP}S.S. Gubser, I.R. Klebanov and A.M. Polyakov, 
{\em Gauge Theory Correlators from Non-Critical String Theory}, Phys. Lett. 
{\bf B428} (1998) 105, hep-th/9802109. 

\bibitem{Witten}E. Witten, {\em Anti De Sitter Space And Holography},
Adv. Theor. Math. Phys. {\bf 2} (1998) 253, hep-th/9802150.

\bibitem{review}O. Aharony, S.S. Gubser, J. Maldacena, H. Ooguri and Y. Oz,
{\em Large N Field Theories, String Theory and Gravity}, hep-th/9905111.

\bibitem{GibbTown}G.W. Gibbons and P.K. Townsend, {\em Vacuum interpolation in supergravity
via super p-branes}, Phys. Rev. Lett. {\bf 71} (1993) 3754, hep-th/9307049.

\bibitem{MinaWarner}J.A. Minahan and N.P. Warner,
{\em Quark Potentials in the Higgs Phase of Large N Supersymmetric Yang-Mills Theories},
JHEP {\bf 06} (1998) 005, hep-th/9805104.

\bibitem{DouglasTaylor}M.R. Douglas and W. Taylor, 
{\em Branes in the bulk of Anti-de Sitter space}, hep-th/9807225.

\bibitem{TseyYank}A.A. Tseytlin and S. Yankielowicz,
{\em Free energy of N=4 super Yang-Mills in Higgs phase 
and non-extremal D3-brane interactions}, 
Nucl. Phys. {\bf B541} (1999) 145, hep-th/9809032.

\bibitem{Wu}Y. Wu, {\em A Note on AdS/SYM Correspondence on the Coulomb Branch},
hep-th/9809055.

\bibitem{KLT}P. Kraus, F. Larsen, S. Trivedi, 
{\em The Coulomb Branch of Gauge Theory from Rotating Branes},
JHEP {\bf 03} (1999) 003, hep-th/9811120.

\bibitem{KlebWitten}I.R. Klebanov and E. Witten,
{\em AdS/CFT Correspondence and Symmetry Breaking}, 
Nucl. Phys. {\bf B556} (1999) 89, hep-th/9905104.

\bibitem{FGPW1}D.Z. Freedman, S.S. Gubser, K. Pilch and N.P. Warner,
{\em Continuous distributions of D3-branes and gauged supergravity},
hep-th/9906194.

\bibitem{BranSfet}A. Brandhuber and K. Sfetsos,
{\em Wilson loops from multicentre and rotating branes, mass gaps and phase structure in
gauge theories}, hep-th/9906201.

\bibitem{ChepRoiban}I. Chepelev and R. Roiban,
{\em A note on correlation functions in $AdS_5/SYM_4$ correspondence on the Coulomb branch},
hep-th/9906224.

\bibitem{GiddRoss}S.B. Giddings and S.F. Ross,
{\em D3-brane shells to black branes on the Coulomb branch},
hep-th/9907204.

\bibitem{CGLP}M. Cvetic, S.S. Gubser, H. Lu and C.N. Pope,
{\em Symmetric Potentials of Gauged Supergravities in Diverse Dimensions and Coulomb
Branch of Gauge Theories}, hep-th/9909121.

\bibitem{RashVisw}R.C.Rashkov and K.S.Viswanathan,
{\em Correlation functions in the Coulomb branch of N=4 SYM from AdS/CFT correspondence}, 
hep-th/9911160.

\bibitem{GHKK}S.S. Gubser, A. Hashimoto, I.R. Klebanov and M. Krasnitz, 
{\em Scalar Absorption and the Breaking of the World Volume Conformal Invariance}, 
Nucl. Phys. {\bf B526} (1998) 393, hep-th/9803023.

\bibitem{GubserHash}S.S. Gubser and A. Hashimoto, 
{\em Exact absorption probabilities for the D3-brane},
Commun. Math. Phys. {\bf 203} (1999) 325, hep-th/9805140 

\bibitem{GPPZ}L. Girardello, M. Petrini, M. Porrati and A. Zaffaroni,
{\em Novel Local CFT and Exact Results on Perturbations of N=4 Super Yang Mills 
from AdS Dynamics}, 
JHEP {\bf 12} (1998) 022, hep-th/9810126;
{\em Confinement and Condensates Without Fine Tuning in Supergravity Duals of Gauge
Theories},
JHEP {\bf 05} (1999) 026, hep-th/9903026;
{\em The Supergravity Dual of N=1 Super Yang-Mills Theory},
hep-th/9909047.

\bibitem{BilalChu}A. Bilal and C. Chu,
{\em D3 Brane(s) in $AdS_5 \times S^5$ and N =4,2,1 SYM},
Nucl. Phys. {\bf B547} (1999) 179, hep-th/9810195.
 
\bibitem{KPW}A. Khavaev, K. Pilch and N.P. Warner,
{\em New Vacua of Gauged N=8 Supergravity},
 hep-th/9812035 

\bibitem{FGPW2}D.Z. Freedman, S.S. Gubser, K. Pilch and N.P. Warner,
{\em Renormalization Group Flows from Holography--Supersymmetry and a c-Theorem},
hep-th/9904017.

\bibitem{Intr}K. Intriligator, {\em Maximally Supersymmetric RG Flows and AdS Duality},
\newline
hep-th/9909082

\bibitem{Gubser}S.S. Gubser, {\em Non-conformal examples of AdS/CFT}, hep-th/9910117.

\bibitem{KlebNekr}I.R. Klebanov and N.A. Nekrasov,
{\em Gravity Duals of Fractional Branes and Logarithmic RG Flow},
hep-th/9911096.

\bibitem{Warner}N. Warner,
{\em Renormalization Group Flows from Five-Dimensional Supergravity},
hep-th/9911240.

\bibitem{Chand}S. Chandrasekhar, {\em The two-centered problem in general relativity:
the scattering of radiation by two extreme Reissner-Nordstr\"om black-holes},
Proc. R. Soc. Lond. {\bf A421} (1989) 227.

\bibitem{MichStro}J. Michelson and A. Strominger, 
{\em Superconformal Multi-Black Hole Quantum Mechanics},
JHEP {\bf 09} (1999) 005, hep-th/9908044.

\bibitem{Kleb}I.R. Klebanov, 
{\em World Volume Approach to Absorption by Non-dilatonic Branes},
Nucl. Phys. {\bf B496} (1997) 231, hep-th/9702076.

\bibitem{Tsey}A.A. Tseytlin, 
{\em On non-abelian generalisation of Born-Infeld action in string theory},
Nucl. Phys. {\bf B501} (1997) 41, hep-th/9701125.

\bibitem{GubserKleb}S.S. Gubser and I.R. Klebanov,
{\em Absorption by Branes and Schwinger Terms in the World Volume Theory},
Phys. Lett. {\bf B413} (1997) 41, hep-th/9708005.

\bibitem{DKPS}M.R. Douglas, D. Kabat, P. Pouliot and S.H. Shenker,
{\em D-branes and Short Distances in String Theory},
Nucl. Phys. {\bf B485} (1997) 85, hep-th/9608024.

\bibitem{LifsMath}G. Lifschytz and S.D. Mathur,
{\em Supersymmetry and Membrane Interactions in M(atrix) Theory},
Nucl. Phys. {\bf B507} (1997) 621, hep-th/9612087.

\bibitem{Mald2}J. Maldacena,
{\em Probing Near Extremal Black Holes with D-branes},
Phys. Rev. {\bf D57} (1998) 3736, hep-th/9705053;
{\em Branes probing black holes},
Nucl. Phys. Proc. Suppl. {\bf 68} (1998) 17, hep-th/9709099. 

\bibitem{ChepTsey}I. Chepelev and A.A. Tseytlin,
{\em Interactions of type IIB D-branes from D-instanton matrix model},
Nucl. Phys. {\bf B511} (1998) 629, hep-th/9705120;
{\em Long-distance interactions of branes: correspondence between supergravity and super
Yang-Mills descriptions},
Nucl. Phys. {\bf B515} (1998) 73, hep-th/9709087; 
A.A. Tseytlin,
{\em Interactions Between Branes and Matrix Theories},
Nucl. Phys. Proc. Suppl. {\bf 68} (1998) 99, hep-th/9709123.

\bibitem{DineSeiberg}M. Dine and N. Seiberg,
{\em Comments on Higher Derivative Operators in Some SUSY Field Theories},
Phys. Lett. {\bf B409} (1997) 239, hep-th/9705057.

\bibitem{LMRS} S. Lee, S. Minwalla, M. Rangamani and N. Seiberg,
{\em Three-Point Functions of Chiral Operators in D=4, ${\cal N}=4$ 
SYM at Large N},
Adv. Theor. Math. Phys. {\bf 2} (1998) 697, hep-th/9806074.

\bibitem{HFS} E. D'Hoker, D.Z. Freedman and W. Skiba,
{\em Field Theory Tests for Correlators in the AdS/CFT Correspondence},
Phys. Rev. {\bf D59} (1999) 045008, hep-th/9807098.

\bibitem{GRKP} F. Gonzalez-Rey, B. Kulik and I.Y. Park,
{\em Non-renormalization of two and three Point Correlators of 
N=4 SYM in N=1 Superspace},
Phys. Lett. {\bf B455} (1999) 164, hep-th/9903094.

\bibitem{IntrSkiba} K. Intriligator,
{\em Bonus Symmetries of N=4 Super-Yang-Mills Correlation Functions via AdS
Duality},
Nucl. Phys. {\bf B551} (1999) 575, hep-th/9811047;
K. Intriligator and W. Skiba,
{\em Bonus Symmetry and the Operator Product Expansion of N=4
Super-Yang-Mills},
Nucl.Phys. {\bf B559} (1999) 165, hep-th/9905020.

\bibitem{EHW} B. Eden, P.S. Howe and P.C. West,
{\em Nilpotent invariants in N=4 SYM},
Phys. Lett. {\bf B463} (1999) 19, hep-th/9905085;
P.S. Howe, C. Schubert, E. Sokatchev and P.C. West,
{\em Explicit construction of nilpotent covariants in N=4 SYM},
hep-th/9910011.

\bibitem{PetkSken} A. Petkou and K. Skenderis,
{\em A non-renormalization theorem for conformal anomalies},
Nucl. Phys. {\bf B561} (1999) 100, hep-th/9906030.

\end{thebibliography}
\end{document}